\documentclass[twocolumn]{aastex631}

\usepackage{hyperref}
\usepackage{amsmath,amstext}
\usepackage[figure,figure*]{hypcap}
\usepackage{amssymb}
\usepackage{nicefrac}
\usepackage{graphicx}

\usepackage{wrapfig}
\usepackage{multirow}
\usepackage{comment}
\usepackage{nicefrac}
\usepackage{array}

\def\sun{$_{\odot}$}

\def\sun{$_{\odot}$}

\def\ergscm2{erg s$^{-1}$ cm$^{-2}$}
\def\yr-1{yr$^{-1}$}

\hypersetup{linkcolor=magenta}

\submitjournal{ApJ}
\received{XX}
\accepted{XX}

\shorttitle{A Host Galaxy Morphology Link Between QPEs and TDEs
} 
\shortauthors{Gilbert {\it et al.}}

\begin{document}

\title{A Host Galaxy Morphology Link Between \\ Quasi-Periodic Eruptions and Tidal Disruption Events}

\correspondingauthor{John Ruan}
\email{jruan@ubishops.ca}

\author{Olivier Gilbert}
\affil{Department of Physics \& Astronomy, Bishop's University, 2600 rue College, Sherbrooke, Québec, J1M 1Z7, Canada}
\affil{Département de Physique, de génie physique et d’optique, Université Laval, Québec, G1V 0A6, Canada}

\author[0000-0001-8665-5523]{John~J.~Ruan}
\affil{Department of Physics \& Astronomy, Bishop's University, 2600 rue College, Sherbrooke, Québec, J1M 1Z7, Canada}

\author[0000-0002-3719-940X]{Michael~Eracleous}
\affil{Department of Astronomy \& Astrophysics and Institute for Gravitation and the Cosmos, Penn State University, 525 Davey Lab, 251 Pollock Road, University Park, PA 16802, USA}

\author[0000-0001-6803-2138]{Daryl Haggard}
\affiliation{McGill Space Institute and Department of Physics, McGill University, 3600 rue University, Montreal, Quebec, H3A 2T8, Canada}

\author[0000-0001-8557-2822]{Jessie~C.~Runnoe}
\affil{Department of Physics and Astronomy, Vanderbilt University, Nashville, TN 37235, USA}

\begin{abstract}
The physical processes that produce X-ray Quasi-Periodic Eruptions (QPEs) recently discovered from the nuclei of several low-redshift galaxies are mysterious. Several pieces of observational evidence strongly suggest a link between QPEs and Tidal Disruption Events (TDE). Previous studies also reveal that the morphologies of TDE host galaxies are highly concentrated, with high Sérsic indicies, bulge-to-total light (\emph{B/T}) ratios, and stellar surface mass densities relative to the broader galaxy population. We use these distinctive properties to test the link between QPEs and TDEs, by comparing these parameters of QPE host galaxies to TDE host galaxies. We employ archival Legacy Survey images of a sample of 9 QPE host galaxies and a sample of 13 TDE host galaxies, and model their surface brightness profiles. We show that QPE host galaxies have high Sérsic indices of $\sim$3, high \emph{B/T} ratios of $\sim$0.5, and high surface mass densities of $\sim$10$^{10}$~$M_\odot$~kpc$^{-2}$. These properties are similar to TDE host galaxies, but are in strong contrast to a mass- and redshift-matched control sample of galaxies. We also find tentative evidence that the central black holes in both QPE and TDE host galaxies are undermassive relative to their stellar mass. The morphological similarities between QPE and TDE host galaxies at the population level add to the mounting evidence of a physical link between these phenomena, and favor QPE models that also invoke TDEs.
\end{abstract}

\section{Introduction}\label{sec:intro}

Quasi-periodic eruptions (QPEs) are X-ray flares emitted from the nuclei of a small handful of low-redshift galaxies, and they display a diverse array of puzzling properties. These quasi-periodic X-ray flares have luminosities of $\sim$10$^{42-44}$~erg~s$^{-1}$, and recurrence times of $\sim$2--50 hours. QPEs have been detected in a total of 9 galaxies to date, through both blind searches in wide-field X-ray surveys such as eROSITA \citep[e.g.,][]{Arcodia21,Arcodia24}, and serendipitous discoveries in targeted X-ray observations \citep[e.g.,][]{Miniutti19, Giustini20, Nicholl24}. Longer-term monitoring of this sample of QPEs has further revealed a variety of puzzling behaviors, such as the observed disappearance of QPEs in the long-term X-ray light curve, and their subsequent reappearance months later with significantly different periods and fluxes \citep[e.g.,][]{miniutti23a, miniutti23b}, or slow fading of the QPE fluxes over years \citep{Pasham23}. Furthermore, although many QPE host galaxies display X-ray emission during quiescence (i.e., between eruptions) and have optical emission line ratio signatures of active galactic nuclei (AGN) in their spectra \citep{Wevers22}, others are
nearly quiescent in X-rays (with Eddington ratios of $L_\mathrm{X}/L_\mathrm{Edd} \sim 10^{-4}$ to $10^{-5}$; \citealt{Arcodia21}), with no spectroscopic signatures of AGN. This rich diversity of properties poses challenges to theoretical models, and additional insights from observations are needed.
    
The physical origin of QPEs is highly uncertain, and many theoretical models have been proposed to explain this phenomenon. Current QPE models can be divided broadly into three categories. First, if QPE host galaxies harbor AGN, instabilities \citep{Sniegowska20} or tears \citep{raj21} in their accretion disks can lead to quasi-periodic outbursts, akin to QPEs \citep{pan22, pan23, kaur23, Sniegowska23}. Second, a secondary star in a eccentric orbit around the central massive black hole (MBH) can be repeatedly stripped at periapsis, resulting in either a partial Tidal Disruption Event \citep[TDE;][]{King20, King22, Metzger22, Wang22, Lu23} or mass transfer through Roche lobe overflow \citep{krolik22, zhao22, linial23a, Wang24}, causing QPEs. Finally, a stellar-mass secondary in an orbit that intersects an existing accretion disk (either from an AGN or a TDE) around the central MBH could cause QPEs, due to collisions as the secondary punches through the disk \citep{xian21, sukova21, linial23b, franchini23, Tagawa23, Yao24, Zhou24}. The possibilities involving a stellar-mass secondary orbiting the MBH are particularly exciting, as they represent the electromagnetic counterparts to Extreme Mass Ratio Inspiral (EMRI) systems that could be detected in gravitational waves by the \emph{Laser Interferometer Space Antenna} \citep[LISA;][]{Amaro-Seoane_2017, Amaro-Seoane_2022}. Periodic electromagnetic flares from these EMRI systems have long been predicted, well before the more recent discovery of QPEs \citep[e.g.,][]{Zalamea10, MacLeod13, MacLeod16, Metzger17, Shen19}.
    
Some pieces of observational evidence intriguingly suggest an link between QPEs and TDEs. For example, the current sample of QPEs all occur in relatively low-mass galaxies (with stellar masses of $M_\star \sim 10^{10}~M_\odot$) that harbor low-mass central MBH \citep{Wevers22}. This preference for low-mass host galaxies is also observed in TDE host galaxies \citep{french16, Law-Smith17, hammerstein21}, for which low-mass MBHs are required for the TDE to occur outside of the MBH event horizon and be observable \citep{Hills75}. Furthermore, optical spectra of QPE host galaxies have revealed post-starburst and quiescent Balmer-strong stellar populations \citep{Wevers22}, reminiscent of TDE host galaxies \citep{french16, Law-Smith17, hammerstein21}.
Recently, \citet{Wevers24a} compared Integral Field Unit spectra of QPE host galaxies to similar observation of TDEs \citep{Wevers24b}, and showed that both TDE and QPE host galaxies display extended emission line regions that indicate a recent ionizination event by a luminous non-stellar continuum. Finally, QPEs have been recently been directly observed in the X-ray light curve of at least one TDE \citep[AT2019qiz;][]{Nicholl24}. This discovery is the strongest and most direct piece of evidence linking TDEs to QPEs, but this evidence has only been observed for one object. Although it is now clear that at least some QPEs occur in TDEs, the extent and nature of the connection between TDEs and QPEs remains an open question.
    
TDE host galaxies are known to have distinct morphological properties \citep{French20}, which would also be observed in QPE host galaxies if these phenomena are linked. Studies using samples of TDEs have revealed that TDE host galaxies are more centrally concentrated, with significantly higher Sérsic indicies relative to both comparison samples of galaxies matched in black hole mass \citep{Law-Smith17}, as well as the broader galaxy population \citep{hammerstein21}. Furthermore, \citet{Graur18} show that TDE host galaxies have higher stellar surface mass densities relative to a comparison sample of the broader galaxy population. These galaxy-scale morphological properties of TDE host galaxies may not be surprising, as it is possible that the high stellar densities in the nuclear regions of the hosts lead to increased rates of close stellar interactions with the central MBH. Regardless of the exact reasons for the high central stellar concentration of TDE host galaxies, we can test the link between TDEs and QPEs by determining whether QPE host galaxies also share these distinctive properties.


\begin{figure*} [t!]
\centering
\includegraphics[scale=0.34,angle=0]{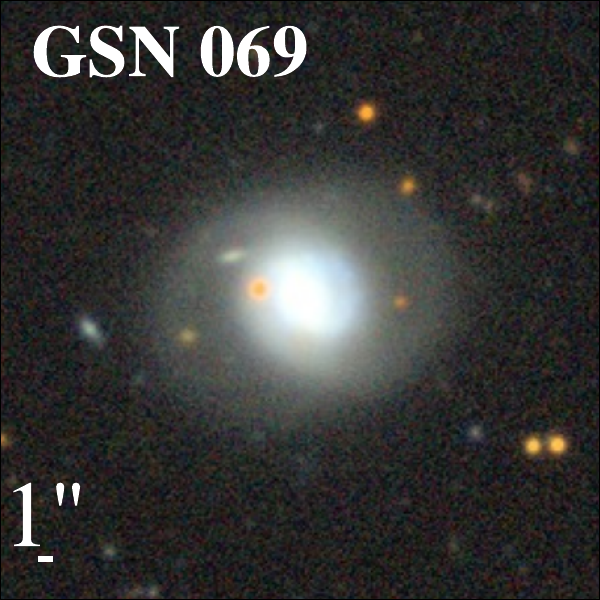}
\includegraphics[scale=0.34,angle=0]{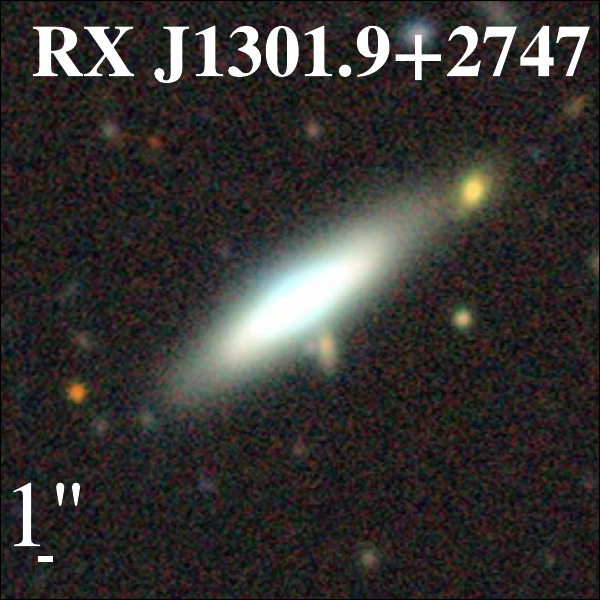}
\includegraphics[scale=0.34,angle=0]{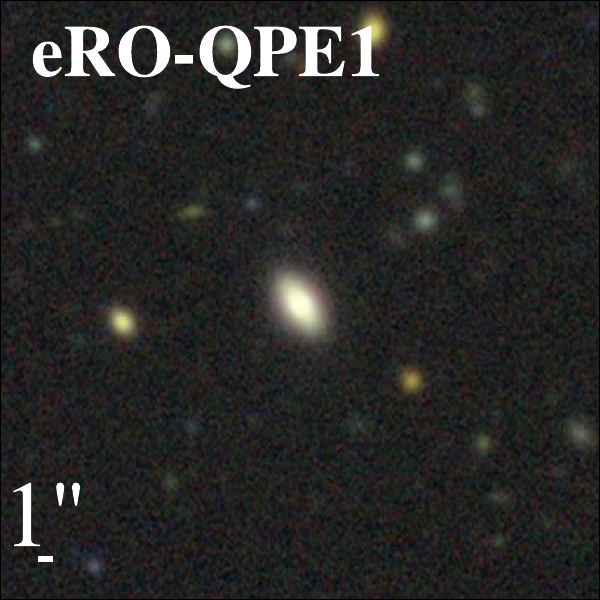}
\includegraphics[scale=0.34,angle=0]{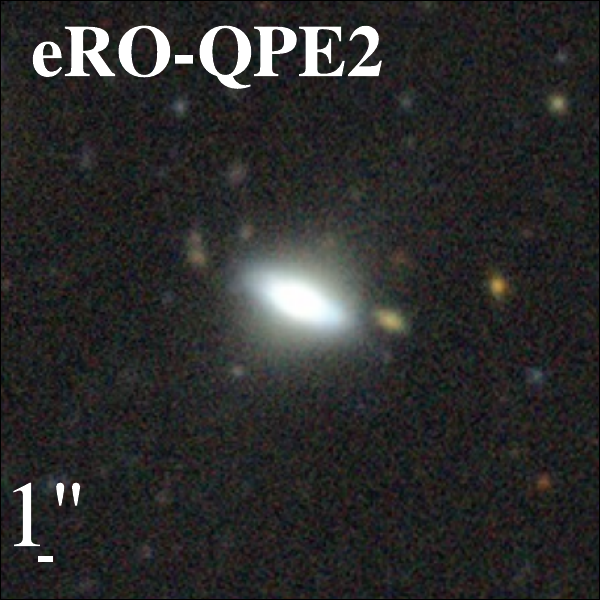}
\includegraphics[scale=0.34,angle=0]{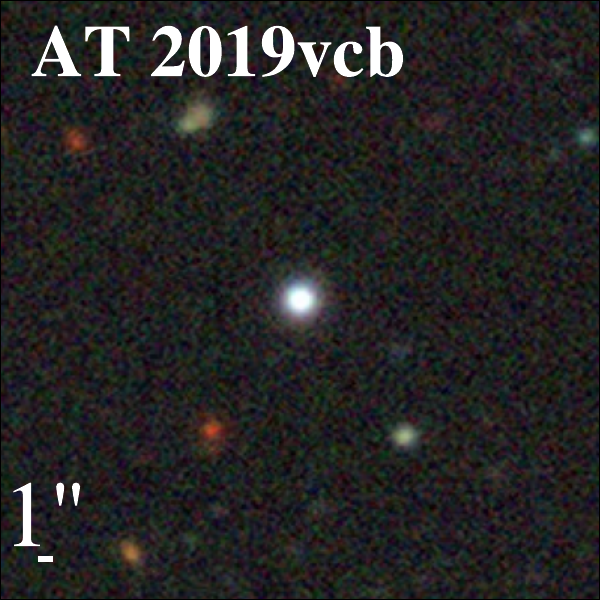}\\
\vspace{2pt}
\includegraphics[scale=0.34,angle=0]{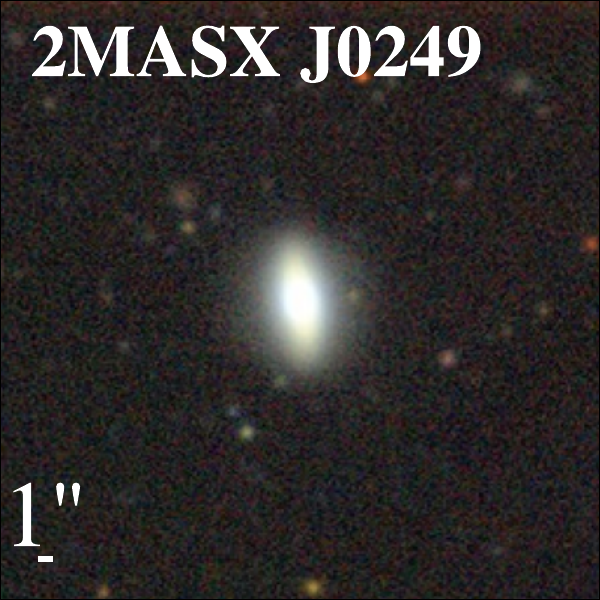}
\includegraphics[scale=0.34,angle=0]{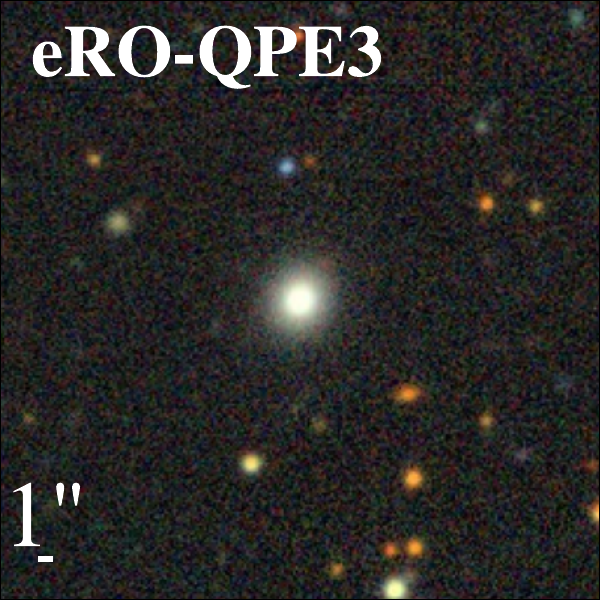}
\includegraphics[scale=0.34,angle=0]{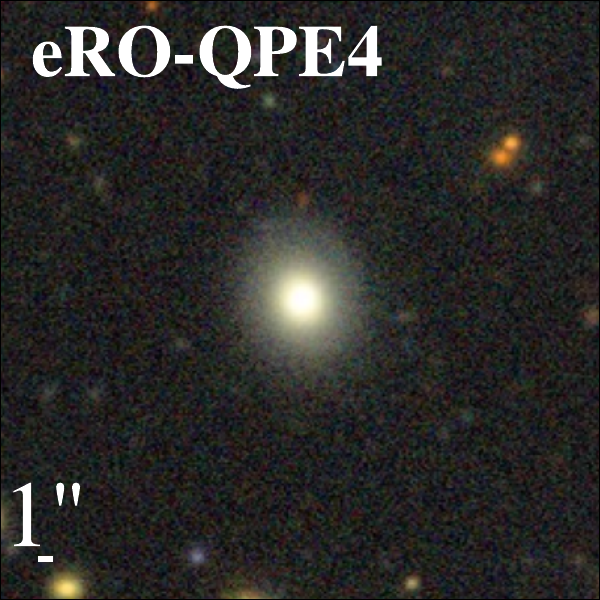}
\includegraphics[scale=0.34,angle=0]{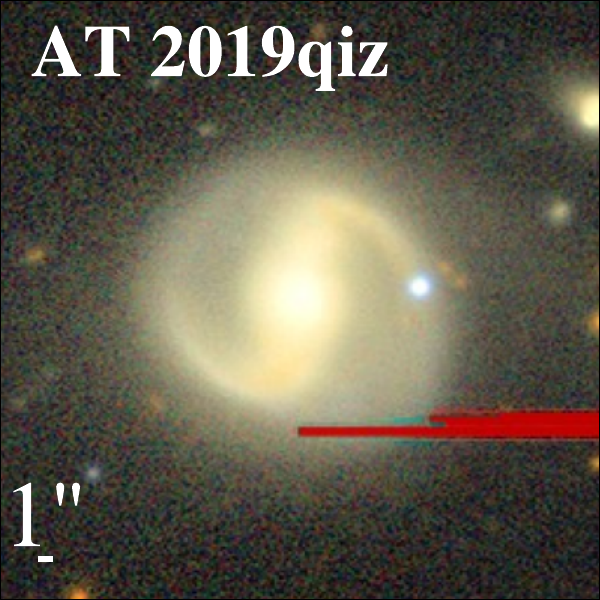}
\figcaption{Legacy Survey \textit{griz} band image cutouts of our sample of 9 QPE host galaxies. These 9 QPE host galaxies include 3 QPE+TDE host galaxies, and are listed in Table~\ref{tab:host_properties}.}
\label{fig:qpe_cutouts}
\end{figure*}
    
In this work, we test the link between TDEs and QPEs, by investigating the morphological properties of QPE host galaxies in archival optical imaging, and comparing them to TDE host galaxies. This comparison requires careful fitting of the surface brightness profiles in imaging, and has not previously been performed. We use archival optical images from the DESI Legacy Imaging Surveys \citep[Legacy Survey;][]{Dey19} of the current sample of 9 QPE host galaxies, along with 13 TDE host galaxies, and model their surface brightness profiles using a uniform approach. We compare the Sérsic indices, bulge-to-total light ratios, and stellar mass densities of the QPE hosts to the TDE hosts, as well as a mass- and redshift-matched control sample of galaxies. We find that QPE host galaxies share similar morphological properties with TDE host galaxies, which are distinct from other galaxies of similar mass and redshift.
    
The outline of this paper is as follows. In Section~\ref{sec:images_modeling}, we describe our QPE and TDE host galaxy samples, archival imaging data, and our surface brightness modeling. In Section~\ref{sec:discussion}, we compare the morphological properties of QPE host galaxies to both TDE host galaxies and the control galaxy sample, and discuss the implications of our results for QPE models. We briefly summarize and conclude in Section~\ref{sec:conclusions}. Throughout the paper, we assume a standard cosmology with $\Omega_\mathrm{m} = 0.31$, $\Omega_\Lambda = 0.69$, and $H_0 = 67$ km s$^{-1}$ Mpc$^{-1}$, consistent with \citet{Planck_Collaboration_2016}.


\begin{figure*} [t!]
\centering
\includegraphics[scale=0.53,angle=0]{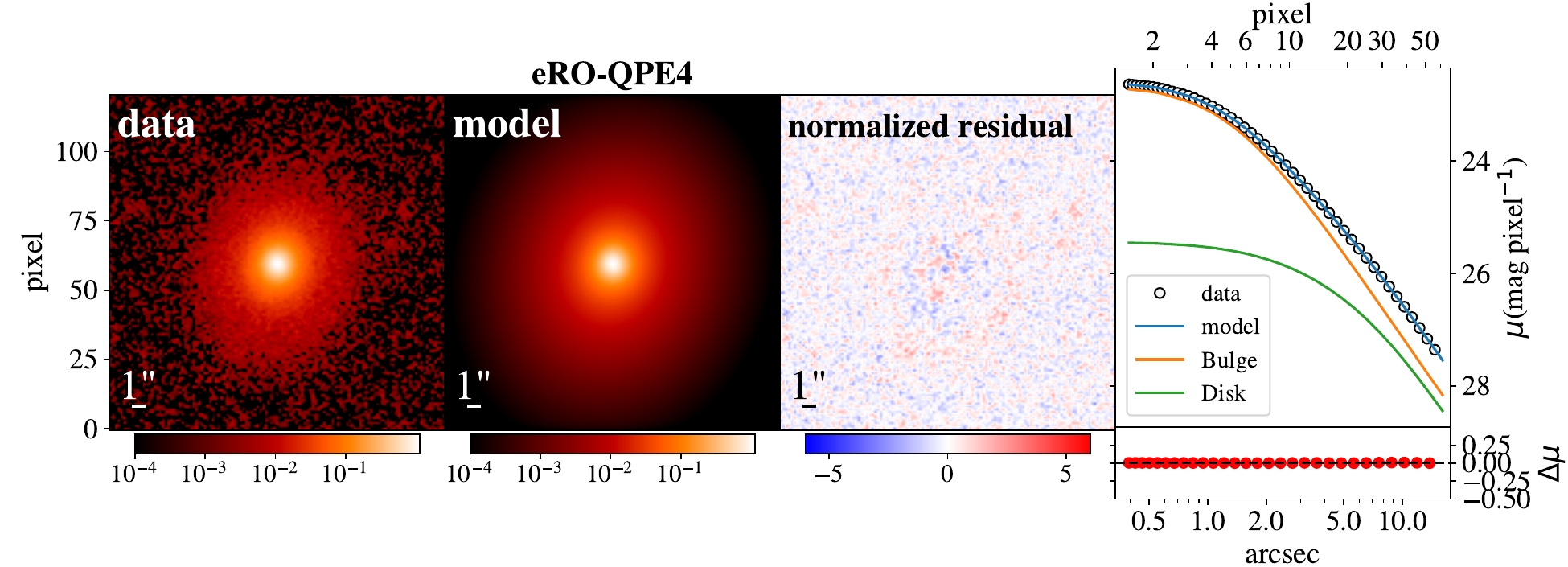} 
\figcaption{Example surface brightness profile bulge+disk fits to the QPE host galaxy eRO-QPE4. The left panel shows the 121$\times$121 pixel \textit{g}-band image cutout centered on the host galaxy. The middle-left panel shows our best-fit two-component (disk+bulge) Sérsic ellipse model. The middle-right panel shows the residual of our model. The right panel shows the one-dimensional surface brightness profile fit, where the observed data is represented by black circles, the bulge by the orange line, the disk by the green line, and the total model (i.e., bulge+disk) by the blue line. The residuals are shown as the red circles under the plot.}
\label{fig:qpe_sersic_fits1}
\end{figure*}

\begin{deluxetable*}{cccccccc}[t!]
\centering
\tablewidth{0pc}
\tabletypesize{\footnotesize}
\tablecaption{Our sample of QPE, QPE+TDE, and TDE host galaxies. Columns include the object name, redshift, half-light radius, Sérsic index, $g$-band bulge-to-total-light ratio, stellar surface mass density, and total stellar mass. Uncertainties are at 1$\sigma$ confidence level. References reporting the discovery of each object are in the footnotes.
\label{tab:host_properties}
}
\tablecolumns{8}
\tablehead{  
\colhead{Object} & \colhead{Redshift} & \colhead{Half-light} & \colhead{Sérsic}  & \colhead{$(B/T)_g$} & \colhead{Stellar Mass}  & \colhead{Stellar Mass} & \colhead{Black Hole} \\ 
\colhead{} & \colhead{$z$} & \colhead{Radius $r_{50}$} & \colhead{Index $n$} & \colhead{ratio}  & \colhead{Density $\log(\Sigma_{M_\star})$} & \colhead{$\log(M_\star)$} & \colhead{Mass $\log(M_\mathrm{BH})$}\\
\colhead{} & \colhead{} & \colhead{[kpc]} & \colhead{} & \colhead{} & \colhead{[$M_\odot$~kpc$^{-2}$]} & \colhead{[$M_\odot$]} & \colhead{[$M_\odot$]}
}
\startdata
\multicolumn{8}{c}{\emph{QPE host galaxies}}\\
\hline
\hline
GSN 069$^{\rm a}$ & $0.018$ & $2.87_{-0.01}^{+0.01}$ & $5.06_{-0.01}^{+0.01}$ & $0.38_{-0.01}^{+0.01}$ & $8.91_{-0.2}^{+0.2}$ & $9.82_{-0.2}^{+0.2}$ & $5.99_{-0.5}^{+0.5}$ \\
\vspace{2pt}
RX J1301.9+2747$^{\rm b}$ & $0.024$ & $2.04_{-0.01}^{+0.01}$ & $2.67_{-0.01}^{+0.01}$ & $0.17_{-0.01}^{+0.01}$ & $9.56_{-0.2}^{+0.2}$ & $10.18_{-0.2}^{+0.2}$ & $6.65_{-0.4}^{+0.4}$ \\
\vspace{2pt}
eRO-QPE1$^{\rm c}$ & $0.05$ & $1.40_{-0.01}^{+0.01}$ & $2.02_{-0.02}^{+0.02}$ & $0.36_{-0.02}^{+0.02}$ & $9.61_{-0.2}^{+0.2}$ & $9.90_{-0.2}^{+0.2}$ & $5.78_{-0.6}^{+0.6}$ \\
\vspace{2pt}
eRO-QPE2$^{\rm c}$ & $0.018$ & $1.09_{-0.01}^{+0.01}$ & $1.18_{-0.01}^{+0.01}$ & $0.39_{-0.01}^{+0.01}$ & $9.35_{-0.2}^{+0.2}$ & $9.43_{-0.2}^{+0.2}$ & $4.96_{-0.5}^{+0.5}$ \\
\vspace{2pt}
eRO-QPE3$^{\rm d}$ & $0.024$ & $0.58_{-0.01}^{+0.01}$ & $2.80_{-0.04}^{+0.04}$ & $0.97_{-0.02}^{+0.02}$ & $9.84_{-0.2}^{+0.2}$ & $9.37_{-0.2}^{+0.2}$ & $6.72_{-0.1}^{+0.3}$ \\
\vspace{2pt}
eRO-QPE4$^{\rm d}$ & $0.044$ & $1.58_{-0.01}^{+0.01}$ & $3.26_{-0.02}^{+0.02}$ & $0.57_{-0.02}^{+0.02}$ & $9.80_{-0.2}^{+0.2}$ & $10.20_{-0.2}^{+0.2}$ & $7.83_{-0.3}^{+0.2}$ \\
\hline
\multicolumn{8}{c}{\emph{TDE+QPE host galaxies}}\\
\hline
\hline
AT 2019vcb$^{\rm e}$ & $0.088$ & $0.84_{-0.01}^{+0.01}$ & $2.03_{-0.08}^{+0.09}$ & $0.53_{-0.03}^{+0.03}$ & $10.01_{-0.3}^{+0.2}$ & $9.86_{-0.3}^{+0.2}$ & $6.81_{-0.1}^{+0.1}$ \\
\vspace{2pt}
2MASX J0249$^{\rm f}$ & $0.019$ & $0.96_{-0.01}^{+0.01}$ & $2.20_{-0.01}^{+0.01}$ & $0.79_{-0.01}^{+0.01}$ & $9.62_{-0.2}^{+0.2}$ & $9.59_{-0.2}^{+0.2}$ & $5.29_{-0.6}^{+0.6}$ \\
\vspace{2pt}
AT 2019qiz$^{\rm g}$ & $0.015$ & $9.75_{-0.01}^{+0.01}$ & $3.74_{-0.01}^{+0.01}$ & $0.54_{-0.01}^{+0.01}$ & $8.33_{-0.2}^{+0.2}$ & $10.31_{-0.2}^{+0.2}$ & $6.18_{-0.4}^{+0.4}$ \\
\hline
\multicolumn{8}{c}{\emph{TDE host galaxies}}\\
\hline
\hline
ASASSN-14ae$^{\rm h}$ & $0.044$ & $2.39_{-0.01}^{+0.01}$ & $1.37_{-0.01}^{+0.01}$ & $0.16_{-0.01}^{+0.01}$ & $9.30_{-0.2}^{+0.2}$ & $10.06_{-0.2}^{+0.2}$ & $5.42_{-0.5}^{+0.5}$ \\
\vspace{2pt}
ASASSN-14li$^{\rm i}$ & $0.021$ & $0.45_{-0.01}^{+0.01}$ & $3.99_{-0.01}^{+0.02}$ & $0.71_{-0.01}^{+0.01}$ & $10.50_{-0.2}^{+0.2}$ & $9.81_{-0.2}^{+0.2}$ & $6.23_{-0.4}^{+0.4}$ \\
\vspace{2pt}
PTF-09ge$^{\rm j}$ & $0.064$ & $3.10_{-0.02}^{+0.02}$ & $1.89_{-0.01}^{+0.01}$ & $0.37_{-0.03}^{+0.03}$ & $9.23_{-0.2}^{+0.2}$ & $10.21_{-0.2}^{+0.2}$ & $6.31_{-0.4}^{+0.4}$ \\
\vspace{2pt}
RBS 1032$^{\rm k}$ & $0.026$ & $0.88_{-0.01}^{+0.01}$ & $1.18_{-0.01}^{+0.01}$ & $0.07_{-0.01}^{+0.01}$ & $9.68_{-0.2}^{+0.2}$ & $9.57_{-0.2}^{+0.2}$ & $5.25_{-0.6}^{+0.7}$ \\
\vspace{2pt}
SDSS J1323$^{\rm l}$ & $0.088$ & $3.36_{-0.03}^{+0.03}$ & $2.11_{-0.03}^{+0.03}$ & $0.13_{-0.02}^{+0.02}$ & $9.52_{-0.2}^{+0.2}$ & $10.57_{-0.2}^{+0.2}$ & $6.15_{-0.5}^{+0.5}$ \\
\vspace{2pt}
SDSS J0748$^{\rm m}$ & $0.062$ & $2.61_{-0.01}^{+0.01}$ & $0.91_{-0.01}^{+0.01}$ & $0.04_{-0.01}^{+0.01}$ & $9.53_{-0.2}^{+0.2}$ & $10.36_{-0.2}^{+0.2}$ & $7.25_{-0.5}^{+0.5}$ \\
\vspace{2pt}
SDSS J1342$^{\rm n}$ & $0.037$ & $1.34_{-0.01}^{+0.01}$ & $2.65_{-0.01}^{+0.01}$ & $0.60_{-0.01}^{+0.01}$ & $9.80_{-0.2}^{+0.2}$ & $10.06_{-0.2}^{+0.2}$ & $6.06_{-0.5}^{+0.5}$ \\
\vspace{2pt}
SDSS J1350$^{\rm n}$ & $0.078$ & $3.43_{-0.03}^{+0.03}$ & $2.87_{-0.03}^{+0.03}$ & $0.22_{-0.01}^{+0.01}$ & $9.31_{-0.2}^{+0.2}$ & $10.38_{-0.2}^{+0.2}$ & $7.47_{-0.6}^{+0.6}$ \\
\vspace{2pt}
SDSS J0952$^{\rm n}$ & $0.079$ & $2.28_{-0.02}^{+0.03}$ & $3.68_{-0.06}^{+0.06}$ & $0.58_{-0.02}^{+0.03}$ & $9.84_{-0.2}^{+0.2}$ & $10.56_{-0.2}^{+0.2}$ & $7.04_{-0.4}^{+0.4}$ \\
\vspace{2pt}
SDSS J1201$^{\rm o}$ & $0.146$ & $3.83_{-0.05}^{+0.05}$ & $3.44_{-0.06}^{+0.07}$ & $0.81_{-0.07}^{+0.06}$ & $9.73_{-0.2}^{+0.2}$ & $10.90_{-0.2}^{+0.2}$ & $7.18_{-0.4}^{+0.4}$ \\
\vspace{2pt}
PTF-15af$^{\rm p}$ & $0.079$ & $2.85_{-0.03}^{+0.03}$ & $2.30_{-0.03}^{+0.04}$ & $0.56_{-0.05}^{+0.05}$ & $9.64_{-0.2}^{+0.2}$ & $10.55_{-0.2}^{+0.2}$ & $6.88_{-0.4}^{+0.4}$ \\
\vspace{2pt}
AT2018dyk$^{\rm q}$ & $0.037$ & $13.28_{-0.05}^{+0.07}$ & $3.49_{-0.01}^{+0.04}$ & $0.45_{-0.01}^{+0.01}$ & $9.38_{-0.1}^{+0.4}$ & $11.62_{-0.1}^{+0.4}$ & $7.00_{-0.4}^{+0.4}$ \\
\vspace{2pt}
ASASSN18zj$^{\rm r}$ & $0.046$ & $1.15_{-0.01}^{+0.01}$ & $1.75_{-0.01}^{+0.01}$ & $0.27_{-0.01}^{+0.01}$ & $9.92_{-0.2}^{+0.2}$ & $10.05_{-0.2}^{+0.2}$ & $5.68_{-0.5}^{+0.5}$ \\
\enddata

\vspace{8pt}
$^{\rm a}$\citet{Miniutti19}; 
$^{\rm b}$\citet{Giustini20};
$^{\rm c}$\citet{Arcodia21};
$^{\rm d}$\citet{Arcodia24};
$^{\rm e}$\citet{Quintin23};
$^{\rm f}$\citet{Chakraborty21};
$^{\rm g}$\citet{Nicholl20};
$^{\rm h}$\citet{Holoien14};
$^{\rm i}$\citet{Jose14};
$^{\rm j}$\citet{Arcavi14};
$^{\rm k}$\citet{Maksym14};
$^{\rm l}$\citet{Read06};
$^{\rm m}$\citet{Wang11};
$^{\rm n}$\citet{Wang12};
$^{\rm o}$\citet{Saxton12}
$^{\rm p}$\citet{Blagorodnova19}
$^{\rm q}$\citet{Huang23}
$^{\rm r}$\citet{Dong18}
\end{deluxetable*}


\section{Modeling of QPE and TDE \ Host Galaxy Images} \label{sec:images_modeling}

\subsection{QPE Host Galaxies Sample}\label{ssc:qpe_sample}

For our QPE host galaxy sample, we use all 9 currently-known QPE hosts in the literature \citep{Miniutti19, Arcodia21, Quintin23, Chakraborty21, Arcodia24}, listed in Table~\ref{tab:host_properties}. This sample of QPEs is inhomogenous, and was discovered through a variety of approaches, including in wide-field X-ray surveys (e.g., eROSITA; \citealt{Predehl03}), and serendipitous discoveries in targeted \emph{XMM-Newton} and \emph{Chandra} observations \citep[e.g.,][]{Giustini20, Miniutti19}. We emphasize that our QPE and TDE host galaxy samples are not mutually exclusive, and include QPE+TDE hosts (i.e., galaxies in which both a TDE and QPEs were observed). Specifically, our sample of 9 QPE host galaxies includes one with {\it confirmed} QPEs in a {\it confirmed} TDE host (AT 2019qiz), and two with {\it candidate} QPEs in {\it confirmed} TDE hosts (AT 2019vcb and 2MASX J0249). All 9 of these QPE host galaxies have archival imaging in the Legacy Survey. 

\subsection{TDE Host Galaxies Sample}\label{ssc:TDE_sample}

For our TDE host galaxies sample, we use the compilation of 13 TDE hosts from \citet{French20} (listed in Table~\ref{tab:host_properties}) that have black hole masses from the literature, along with the three QPE+TDE hosts. This sample consists primarily of TDEs from \citet{Law-Smith17}, with a few more recent discoveries. Both \citet{French20} and \citet{Law-Smith17} show that these TDE hosts have high Sérsic indicies relative to a comparison sample of galaxies.

\subsection{Imaging Data}\label{ssc:images}

We use archival optical images of the QPE and TDE host galaxies from the Legacy Survey Data Release 10.1\footnote{https://www.legacysurvey.org/}. The Legacy Survey is a wide field imaging survey that combines the Mayall $z$-band Legacy Survey (MzLS), the Dark Energy Camera Legacy Survey (DECaLS), and the Beijing-Arizona Sky Survey (BASS), and is supplemented by additional Dark Energy Camera imaging, covering a total area of $\sim20,000~\mathrm{deg^2}$. We use coadded images produced by the Legacy Survey team, which have median $5\sigma$ AB magnitude depths of $g\sim25.0$, $r\sim24.5$, $i\sim24.2$, $z\sim23.5$. These coadded images include coadd point spread functions (PSFs), which have a typical FWHM of $\sim$1\farcs2. We refer the reader to \citet{Dey19} for more details on the Legacy Survey. Figure~\ref{fig:qpe_cutouts} shows cutouts of the Legacy Survey coadded images for our sample of 9 QPE host galaxies (including the 3 QPE+TDE hosts).

\subsection{Surface Brightness Profile Modeling}\label{ssc:modeling}

We use the \texttt{galight} software \citep{ding20} to model the surface brightness profile of each QPE and TDE host galaxy. We fetch calibrated coadded \textit{griz} images of each host galaxy from the Legacy Survey archive, along with their corresponding weight maps, which we use to calculate noise maps for our fits. We also fetch the coadd PSF at the location of each galaxy, which we use in our modeling. We use \texttt{galight} to first perform source detection in each image, and mask out all sources except for the target host galaxy. We fit the background level in the masked image, and subtract the median value. We use a 121$\times$121 pixel cutout centered on each host galaxy for our detailed fits; in our tests, our fitting results are robust to the choice of this cutout size.
    
We first fit the \textit{r}-band surface brightness profile of each QPE and TDE host galaxy to a single Sérsic model, to obtain a global Sérsic index $n$. Specifically, we use \texttt{galight} to model each host galaxy with a Sérsic ellipse model from \texttt{lenstronomy} \citep{Birrer_2018, Birrer2021}, convolved with the PSF. We use \texttt{emcee} \citep{emcee13} to perform Markov Chain Monte Carlo (MCMC) to produce posterior distributions of the free parameters, including the Sérsic index $n$, and the Sérsic half-light radius $r_{50}$. For the MCMC, we use 120 walkers with 5000 total iterations, including 500 burn-in iterations. The best-fitting values (median, along with the 16th--84th percentiles of the posterior as the uncertainty) of $n$ and $r_{50}$ for each host galaxy are listed in Table~\ref{tab:host_properties}.

We next fit a two-component disk+bulge model to each galaxy, to obtain a $g$-band bulge-to-total light ratio ($B/T$)$_g$. We model the host galaxy with two Sérsic ellipse models from \texttt{lenstronomy}, to represent a disk and a bulge. We refer to these disk and bulge Sérsic indices as $n_d$ and $n_b$, respectively. For some galaxies, these disk+bulge fits produce unphysical results, because the galaxy is either strongly disk-dominated or strongly bulge-dominated, leaving the other component poorly-constrained. Thus, we use the results of our previous single-Sérsic fit to inform our disk+bulge fits. Specifically, for galaxies with $n<1.5$ from our single-Sersic fits (strongly disk-dominated), we fix the bulge-component Sersic index in our bulge+disk fit to $n_b=4$, while leaving $n_d$ free. This ensures that the bulge component does not also fit the disk. Conversely, if $n>3$ from our single-Sersic fits (strongly bulge-dominated), we fix the disk-component Sersic index in our bulge+disk fit to $n_d=1$, while leaving $n_b$ to be free. Finally, in the intermediate case where $1.5\leq n \leq 3$ from our single-Sersic fits (both disk and bulge components are relatively equal), we leave both $n_d$ and $n_b$ free. We produce posterior distributions for the free parameters by performing MCMC, similar to our single-Sérsic fits above. The resultant ($B/T$)$_g$ ratios are listed in Table~\ref{tab:host_properties}. We show an example of our disk+bulge fit to a QPE host galaxy from our sample (eRO-QPE4) in Figure~\ref{fig:qpe_sersic_fits1}.

There is also spectroscopic evidence for low-luminosity AGN in most QPE host galaxies \citep{Wevers22}, and thus we perform tests of our surface brightness fitting that include an additional point source to represent the AGN. Specifically, we compare our fits using the two-component disk+bulge model described above to those using a three-component disk+bulge+AGN model, where the AGN is a central point source based on the coadd PSF. We find that the two-component disk+bulge model is preferred for all objects, based on its Bayesian Information Criterion (BIC). Thus, we find no evidence for spatially-unresolved AGN emission in any of our QPE and TDE host galaxy images.

\begin{figure} [t!]
\centering
\includegraphics[scale=0.57,angle=0]{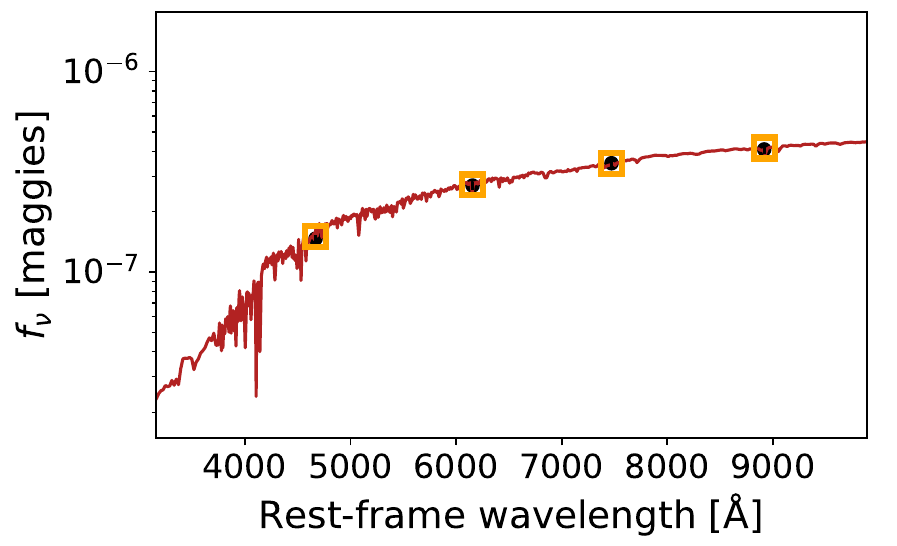} 
\figcaption{Example SED fit from \texttt{prospector} \citep{Johnson21} for the QPE host galaxy eRO-QPE4. The black circles represent the \textit{griz}-band flux densities calculated from our bulge+disk decompositions, the orange squares represent the best-fitting flux densities for each of the four bands, and the red line represents the best-fit SED. From these fits, we obtain stellar masses for the host galaxies.}
\label{fig:qpe_sed}
\end{figure}

We use our two-component disk+bulge fits to compute stellar surface mass densities $\Sigma_\star$ for each host galaxy. We calculate $\Sigma_\star$ using the definition from \citet{Graur18}
    \begin{equation}
    \Sigma_{M_\star} = \frac{M_\star}{r_{50}^2} \quad [M_\odot~\mathrm{kpc}^{-2}],
    \label{eq:density}
    \end{equation}

where $r_{50}$ is the \textit{r}-band half-light radius obtained from our single-Sérsic fits to the surface brightness profile, and $M_\star$ is the total stellar mass of the galaxy. We obtain $M_\star$ through fitting the optical Spectral Energy Distribution (SED) of each host galaxy using our measured \textit{griz}-band apparent magnitudes. To calculate these apparent magnitudes, we add the flux of the fitted bulge and disk components from our bulge+disk decomposition of each galaxy, and correct for Galactic extinction using the dust maps of \citet{Schlafly11}, assuming the $R_V=3.1$ extinction law from \citet{Fitzpatrick99}. We fit the resultant SEDs using the \texttt{prospector} software to obtain $M_\star$. We perform this inference using nested sampling with \texttt{dynesty} \citep{Speagle20}, fixing the redshift to the value from the Sloan Digital Sky Survey \citep[SDSS;][]{York00}, and leaving the following free parameters: stellar mass $M_\star$, metallicity, optical depth at 5500\AA{} (to model the host galaxy dust), the age of the stellar population $t_\mathrm{age}$, and the $e$-folding time $\tau$, such that the star formation history is $\mathrm{SFH}\propto e^{-t_\mathrm{age}/\tau}$ \citep{Bruzual83,Papovich01}. An example of the SED fitting for the QPE host galaxy eRO-QPE4 is shown in Figure~\ref{fig:qpe_sed}. From these fits, we obtain posterior distributions of the free parameters, including $M_\star$, and use the median $M_\star$ value and its 16th--84th percentiles as our uncertainty to calculate $\Sigma_{M_\star}$ in Equation~\ref{eq:density}. The resultant $\Sigma_{M_\star}$ and $M_\star$ for each host galaxy are listed in Table~\ref{tab:host_properties}.

\subsection{Central Massive Black Hole Masses}\label{ssc:MBHmass}
\begin{figure*}
    \centering
    \includegraphics[scale=0.57]{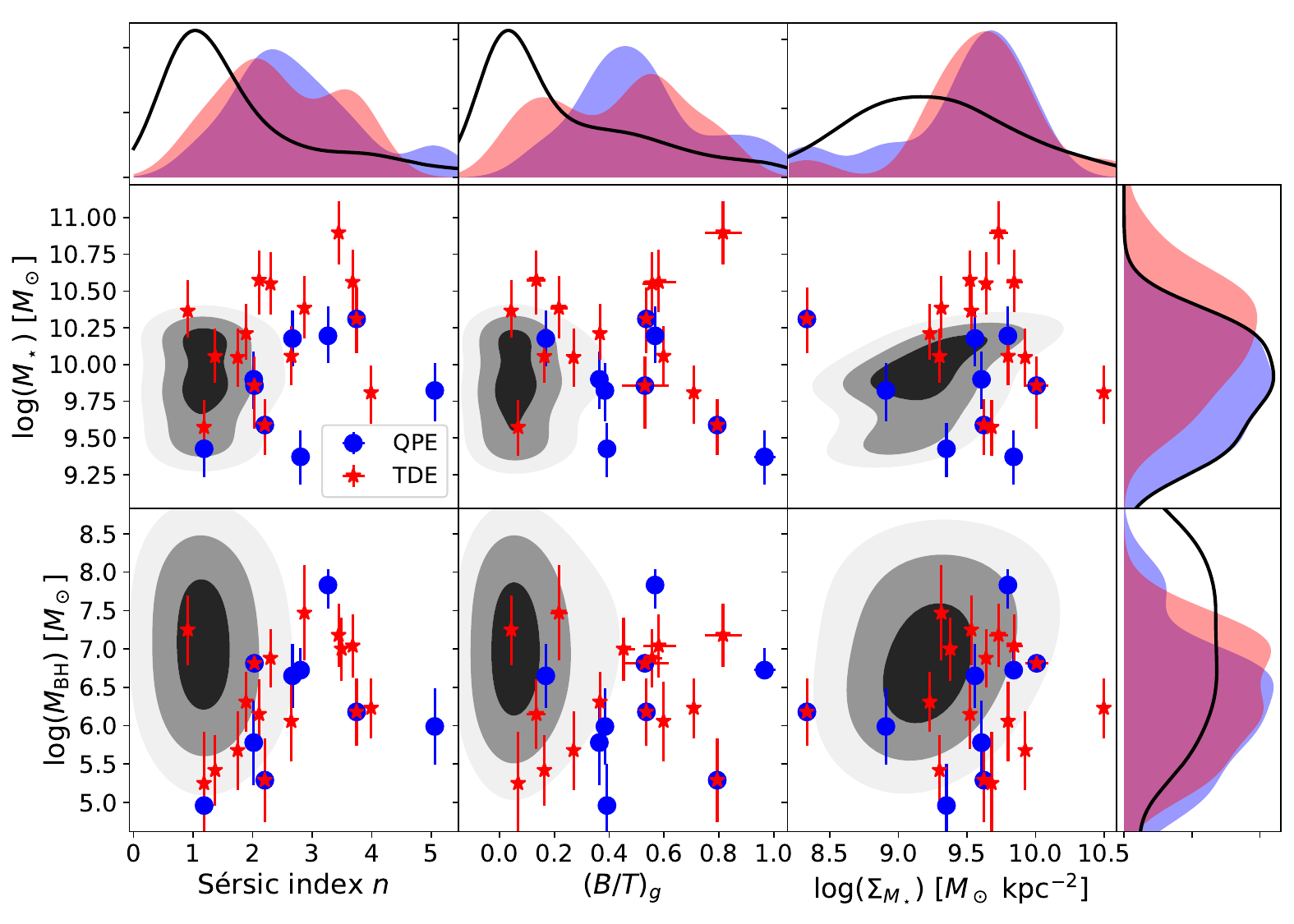}
    \caption{A comparison of our measured morphological properties of QPEs (blue circles) to both TDEs (red stars), and the control galaxy sample (grey contours enclosing 10\%, 30\%, and 50\% of the sample). This control galaxy sample is matched in both stellar mass and redshift to the QPE host galaxy sample. The columns show the distributions in Sérsic index $n$ (left column), $g$-band bulge-to-total light ratio $(B/T)_g$ (center column), and stellar mass density $\Sigma_{M_\star}$ (right column), as a function of galaxy stellar mass $M_\star$ (top row) and black hole mass $M_\mathrm{BH}$. The three objects with a red star in a blue circle are the three QPE+TDE host galaxies. Smoothed histograms of these quantities from a kernel density estimator are also shown. Both QPE and TDE host galaxies have similar high $n$, $(B/T)_g$, $\Sigma_{M_\star}$ relative to the control sample, suggesting a link between QPEs and TDEs based on their host galaxy morphologies. Furthermore, both QPE and TDE hosts have undermassive $M_\mathrm{BH}$ relative to the control sample, suggesting they lie systematically below the $M_\star$-$M_\mathrm{BH}$ galaxy scaling relation.}
    \label{fig:parameter_distributions}
\end{figure*}

We also compare the MBH masses, $M_\mathrm{BH}$, of QPE host galaxies to TDE hosts. Although previous studies have pointed out that QPE host galaxies seem to have low $M_\mathrm{BH}$ that are similar to TDE hosts \citep[e.g.,][]{Wevers22}, no explicit comparisons have been made. We thus estimate $M_\mathrm{BH}$ for each host galaxy based on its empirical relation with velocity dispersion from \citet{Kormendy13}
    \begin{equation}
        \frac{M_{BH}}{10^9 M_\odot} = \left(0.309^{+0.037}_{-0.033}\right)\left(\frac{\sigma_v}{200~\mathrm{km~s^{-1}}}\right)^{4.38\pm0.29},
        \label{eq:mBH_velocityDisp}
    \end{equation}
where $\sigma_v$ is the velocity dispersion in $\mathrm{km~s^{-1}}$, which is calculated from the measured velocity dispersion within the aperture $\sigma_{ap}$ as listed in the SDSS MPA-JHU Galaxy Properties \footnote{https://www.sdss4.org/dr17/spectro/galaxy\textunderscore mpajhu/} catalog. We use the equation 
    \begin{equation}
        \log\left(\frac{\sigma_{ap}}{\sigma_v}\right) = -0.065\log\left(\frac{r_{ap}}{r_{50}}\right)-0.013\left[\log\left(\frac{r_{ap}}{r_{50}}\right)\right]^2
        \label{velocityDispersionRelation}
    \end{equation}
to convert from $\sigma_{ap}$ to $\sigma_v$ \citep{Jorgensen95}, where $r_{50}$ is the $r$-band half-light radius from the \citet{Simard11} catalog, and $r_{ap}$ is the aperture radius of the SDSS observations used in the {MPA-JHU} catalog, which corresponds to $1\farcs5$. The resultant $M_\mathrm{BH}$ for each host galaxy is listed in Table~\ref{tab:host_properties}, and their distributions are shown in Figure~\ref{fig:parameter_distributions}.
    
\section{Discussion}\label{sec:discussion}

\subsection{Comparison of QPE Host Galaxies \\ to TDE Host Galaxies}\label{ssc:qpe_tde_comparison}

We compare our measured parameters for QPE host galaxies to TDE host galaxies, and find that they have similar Sérsic indices $n$, bulge-to-total light ratios $(B/T)_g$, and stellar surface mass densities $\Sigma_{M_\star}$. Specifically, we compare 
our sample of 9 QPE host galaxies (including the 3 QPE+TDE hosts) to our sample of 13 TDE host galaxies (also including the three QPE+TDE hosts). The TDE+QPE hosts are thus included in both the QPE and TDE samples. Figure~\ref{fig:parameter_distributions} compares the distributions of $n$, $(B/T)_g$, $\Sigma_{M_\star}$, $M_\star$, and $M_\mathrm{BH}$, for QPE (blue) and TDE (red) host galaxies. Qualitatively, the distributions of the Sérsic $n$, $(B/T)_g$, $\Sigma_{M_\star}$, $M_\star$, and $M_\mathrm{BH}$ are similar between the QPE and TDE host galaxies. 

We next perform statistical tests to quantitatively demonstrate that the TDE and QPE host galaxy parameter distributions are similar. We first use a 2-sample Kolmogorov-Smirnov (K-S) test to compare the shape of the parameter distributions between TDE and QPE host galaxies. We find that we cannot reject the null hypothesis that the two samples belong to the same distribution for $n$, $(B/T)_g$, $\Sigma_{M_\star}$, $M_\star$, and $M_\mathrm{BH}$, with $p$-values of 0.93, 0.71, 0.98, 0.14 and 0.93, respectively. We next perform an Anderson-Darling test to further compare the tails of the parameter distributions, and also find that $n$, $(B/T)_g$, $\Sigma_{M_\star}$, $M_\star$, and $M_\mathrm{BH}$ all appear to be drawn from the same distribution, with $p$-values of 0.87, 0.61, 0.97, 0.06, and 0.61, respectively. We note that although our QPE and TDE samples are not mutually exclusive, these results are not strongly biased by inclusion of the three QPE+TDE host galaxies. We find that excluding these QPE+TDE hosts in our statistical tests leads to similar conclusions.
    


\subsection{Comparison of QPE Host Galaxies \\ to the Control Sample}\label{ssc:qpe_galaxypopulation_comparison}

To determine whether the properties of QPE host galaxies are distinct, we construct a control sample of galaxies from SDSS, matched in redshift and stellar mass to our QPE hosts. To enable this comparison and place this control sample on Figure~\ref{fig:parameter_distributions}, we build it using various SDSS galaxy catalogs that contain $z$, $n$, $r_{50}$, $(B/T)_g$, $\Sigma_{M_\star}$, $M_\star$ and $M_\mathrm{BH}$. Specifically, we use the \citet{Simard11} catalog of galaxy properties from SDSS imaging to obtain $z$, $n$, $r_{50}$ and $(B/T)_g$, and we obtain $M_\star$ from the \citet{Mendel14} catalog. For each galaxy, we estimate $M_\mathrm{BH}$ using the procedure described in Section~\ref{ssc:MBHmass}.
    
We perform selection cuts on our control sample, so that both its redshift and stellar mass distribution will match our QPE host galaxies. We first exclude galaxies with a redshift $z<0.01$ to prevent aperture bias where the galaxy has a much larger angular size than the SDSS fiber diameter, which could cause the velocity dispersion (and consequently $M_\mathrm{BH}$) to be overestimated. To match our control galaxy sample in mass and redshift to the QPE hosts, we bin our QPE hosts and our control galaxies to a 20$\times$20 grid of $M_\star$ and $z$. We then pick 5,000 control galaxies from each bin containing a QPE host, thus selecting a total of 45,000 galaxies with similar $M_\star$ and $z$ as our QPE hosts to generate our mass- and redshift-matched control sample.

We find that QPE host galaxies have higher $n$, $(B/T)_g$, and $\Sigma_{M_\star}$ in comparison to the mass- and redshift-matched control sample, similar to TDE hosts. Figure~\ref{fig:parameter_distributions} also compares these parameters between the QPE hosts to the control sample, which qualitatively shows a clear difference between these two galaxy samples. To quantify these differences, we calculate the percentage of galaxies above fiducal parameter values for both the QPE and the control sample distributions. We find that for QPE hosts, 89\% of our sample has $n>2$, 89\% has $(B/T)_g>0.35$, 67\% has $\log(\Sigma_{M_\star})>9.5$, and 89\% has $\log(M_\mathrm{BH})<7$, in stark contrast (by a factor of $\sim$2) to 50\%, 41\%, 33\%, and 58\% for our control galaxies, respectively. Figure~\ref{fig:parameter_distributions} also suggests 
that while QPE and TDE hosts have similar MBH masses of $M_\mathrm{BH} \sim 10^{6.5}~M_\odot$, these masses are systematically smaller than the control sample matched in stellar mass and redshift. A 2-sample K-S test gives $p$-values of 0.09, 0.02, 0.11, and 0.16, and an Anderson-Darling test gives $p$-values of 0.25, 0.041, 0.20 and 0.39, for $n$, $(B/T)_g$, $\Sigma_{M_\star}$, and $M_\mathrm{BH}$, respectively. While the statistical significance threshold of $p < 0.05$ is only formally met for $(B/T)_g$, the other properties all exhibit low $p$-values, pointing towards a difference between the QPE and control galaxy samples. Thus, the K-S tests indicate that the  $(B/T)_g$ distribution of QPE host galaxies is distinct from the control sample, while the Anderson-Darling test indicates that the $M_\mathrm{BH}$ distribution of QPE host galaxies is distinct from the control sample at a statistically significant level. We note that the distributions of $M_\star$ for QPE hosts and the control galaxy sample are similar by design (with $p$-value of 1.00), since the control sample is matched in $M_\star$ to the QPE hosts. Although neither the K-S test nor the Anderson-Darling test are able to provide statically robust conclusions for the majority of the parameters, these 
tests are dependent on sample size. Thus, more conclusive statistical tests will have to await the discovery of larger samples of QPEs.

\subsection{ Implications for the Origins of QPEs}\label{ssc:interpretation}

Our results suggest that QPEs and TDEs both result from processes related to an increased rate of close stellar interactions with the central MBH. The similarly high $n$, $(B/T)_g$, and $\Sigma_{M_\star}$ of both QPE and TDE host galaxies relative to the control sample suggest that their host galaxies are bulge-dominated, and have unusually high central stellar concentrations. In these dense stellar environments, the rates of close stellar encounters with the central MBH are elevated, and this scenario has previously been invoked to explain the distinct properties of TDE host galaxies \citep[e.g.,][]{Law-Smith17, Graur18, hammerstein21}. Our results here further extend this explanation to QPEs, which may also owe to close stellar encounters with the central MBH.

The link between QPE and TDE host galaxies supports QPE models that also invoke TDEs, and suggests that these two phenomena may be different manifestations of the same underlying physical process. For example, several studies have suggested that partial TDEs from a star on an elliptical orbit around a MBH could cause QPEs \citep{King20, King22, Metzger22, Wang22, Lu23}, although the exact emission mechanisms are still unclear, even for TDEs. Similarly, a link between QPEs and TDEs would also be consistent with models that invoke a star on an inclined orbit colliding with an accretion disk surrounding the central MBH \citep{xian21, sukova21, linial23b, franchini23, Tagawa23, Yao24, Zhou24}, if this accretion disk is specifically due to a TDE. However, it is still unclear if the link between TDEs and QPEs is causal (i.e., if TDEs are necessarily required for QPEs to occur, or vice versa). More detailed comparisons between simulations and QPE light curves \citep[e.g.,][]{franchini23} will be needed to differentiate between these models, and additional observations of both TDEs and QPEs will clarify their link.

We also find that both QPE and TDE host galaxies may have systematically undermassive black holes, which indirectly also  supports a link between TDEs and QPEs. Figure~\ref{fig:parameter_distributions} shows that the $M_\mathrm{BH}$ distribution of QPE and TDE host galaxies may be systematically smaller than the mass- and redshift-matched control galaxy sample. Since this control sample is matched in stellar mass to our QPE sample, this implies that both QPEs and TDEs may lie below the $M_\mathrm{BH} - M_\star$ relation for galaxies. Previous observational studies have suggested that TDE host galaxies may preferentially host undermassive black holes relative to their host galaxy mass \citep{Ramsden22}. This may owe to a bias in preferentially observing TDEs in galaxies with undermassive $M_\mathrm{BH}$, since the tidal radius for these undermassive black holes is more likely to be outside the event horizon. If QPEs are linked to TDEs, the $M_\mathrm{BH}$ in QPE host galaxies would thus naturally also be undermassive. Thus, our finding of undermassive $M_\mathrm{BH}$ in both TDEs and QPEs also supports this link.

We perform tests to verify that our finding of undermassive $M_\mathrm{BH}$ in both TDE and QPE hosts does not arise from systematic differences in $M_\star$ estimated from different methods and data. Specifically, the $M_\star$ we estimate for TDE and QPE hosts is from the \texttt{prospector} software using Legacy Survey images, whereas the literature $M_\star$ we use for the control sample is from the \citet{Mendel14} catalog based on SDSS imaging. A systematic difference in the $M_\star$ from these two approaches could, in turn, cause a systematic difference in the $M_\mathrm{BH}$ distributions, since our control galaxy sample is matched in $M_\star$ to the QPE hosts. In other words, if our $M_\star$ estimates for QPE hosts from \texttt{prospector} are systematically larger than the $M_\star$ estimates for the control galaxies sample from \citet{Mendel14}, and our control sample is $M_\star$-matched to the QPE hosts, then the QPE hosts would naturally have undermassive $M_\mathrm{BH}$ relative to the control sample. To evaluate this possibility, we use the seven TDE and QPE hosts in our samples that also have $M_\star$ from the \citet{Mendel14} catalog, and directly compare their $M_\star$ values. We find that our $M_\star$ estimates from \texttt{prospector} are actually systematically $\lesssim$0.2~dex \emph{lower} than the values listed in \citet{Mendel14}. This systematic difference is not only small, but carefully correcting for this effect would cause the $M_\mathrm{BH}$ distributions for QPE and TDE hosts in Figure~\ref{fig:parameter_distributions} to become even more undermassive. Thus, our finding of undermassive $M_\mathrm{BH}$ for QPE and TDE hosts cannot owe to systematics stemming from the imaging data or $M_\star$ estimates.

\section{Conclusions}\label{sec:conclusions}
We tested for a possible link between QPEs and TDEs,  based on their host galaxy morphological properties. Using archival Legacy Survey optical images of a sample of QPE and TDE host galaxies, we fitted their surface brightness profiles and measured key morphological parameters in a systematic analysis. We compared these QPE host galaxy parameters to those of TDEs, as well as a control sample of galaxies matched in redshift and stellar mass. Our main findings are:

\begin{enumerate}
    \item The morphological properties of QPE host galaxies are similar to those of TDE hosts, while being distinct from the control galaxy sample. Both QPE and TDE host galaxies not only have similar distributions in Sérsic $n$, bulge-to-total light ratio $(B/T)_g$, and stellar surface mass density $\Sigma_{M_\star}$, but these parameters are also systematically higher than in a mass- and redshift-matched control sample of galaxies. We also tentatively find that both QPE and TDE host galaxies have undermassive black holes compared to the control galaxy sample, which implies that they systematically lie below the $M_\mathrm{BH} - M_\star$ relation for galaxies. 
    
    \item Our results suggest that QPEs and TDEs are linked, and favor QPE models that also invoke TDEs. Specifically, the similarly high $n$, $(B/T)_g$, and $\Sigma_{M_\star}$ of QPE and TDE host galaxies relative to the control sample suggests that both phenomena occur in galaxies with high central stellar concentrations, where the rate of close stellar encounters with the central MBH is higher. This scenario is consistent with QPE models such as repeated partial TDEs, or a stellar-mass secondary orbiting the central MBH that is interacting with a TDE accretion disk. Finally, the undermassive central black holes in both QPE and TDE host galaxies may owe to the upper limit on $M_\mathrm{BH}$ for TDEs to be observable. If QPEs and TDEs are different manifestations of the same underlying phenomenon, this would would naturally explain why MBHs in both QPE and TDE host galaxies are undermassive.
    
\end{enumerate}

The link between QPE and TDEs suggested by our results has implications for multi-messenger observations of EMRI systems detected in gravitational waves. Although using EMRIs to trace orbits in the strong gravity around MBHs is a key science goal of future mHZ gravitational wave experiments such as LISA, it is unclear whether the electromagnetic counterparts to these systems will be detectable. If QPEs are signposts for EMRIs and occur in TDEs, wide-field imaging surveys such as the Legacy Survey of Space and Time \citep[LSST;][]{Ivezic19} on the Rubin Observatory can be used to identify the TDE counterparts to EMRIs. These identifications will enable X-ray follow-up to search for QPEs, and perform multi-messenger science with these systems. Even before the LISA era, long-term X-ray follow-up of more TDEs can unveil additional insights into the exact physical processes that produce QPEs.


\begin{acknowledgments}
J.J.R. thanks Eric Agol for helpful discussions. O.G. acknowledges support from the NSERC Undergraduate Student Research Award program. J.J.R.\ and D.H.\ acknowledge support from the Canada Research Chairs (CRC) program, the NSERC Discovery Grant program, and the NSERC Alliance International program. J.J.R.\ acknowledges funding from the Canada Foundation for Innovation (CFI), and the Qu\'{e}bec Ministère de l’\'{E}conomie et de l’Innovation.

The DESI Legacy Imaging Surveys consist of three individual and complementary projects: the Dark Energy Camera Legacy Survey (DECaLS), the Beijing-Arizona Sky Survey (BASS), and the Mayall z-band Legacy Survey (MzLS). DECaLS, BASS and MzLS together include data obtained, respectively, at the Blanco telescope, Cerro Tololo Inter-American Observatory, NSF’s NOIRLab; the Bok telescope, Steward Observatory, University of Arizona; and the Mayall telescope, Kitt Peak National Observatory, NOIRLab. NOIRLab is operated by the Association of Universities for Research in Astronomy (AURA) under a cooperative agreement with the National Science Foundation. Pipeline processing and analyses of the data were supported by NOIRLab and the Lawrence Berkeley National Laboratory (LBNL). Legacy Surveys also uses data products from the Near-Earth Object Wide-field Infrared Survey Explorer (NEOWISE), a project of the Jet Propulsion Laboratory/California Institute of Technology, funded by the National Aeronautics and Space Administration. Legacy Surveys was supported by: the Director, Office of Science, Office of High Energy Physics of the U.S. Department of Energy; the National Energy Research Scientific Computing Center, a DOE Office of Science User Facility; the U.S. National Science Foundation, Division of Astronomical Sciences; the National Astronomical Observatories of China, the Chinese Academy of Sciences and the Chinese National Natural Science Foundation. LBNL is managed by the Regents of the University of California under contract to the U.S. Department of Energy. The complete acknowledgments can be found at \href{https://www.legacysurvey.org/acknowledgment/}{https://www.legacysurvey.org/acknowledgment/}.

\end{acknowledgments}


\software{
\href{><://docs.astropy.org}{\texttt{astropy}} (\citealt{astropy18}); 
\href{https://github.com/lenstronomy/lenstronomy}{\texttt{lenstronomy}} (\citealt{Birrer_2018, Birrer2021});
\href{https://galight.readthedocs.io}{\texttt{galight}} (\citealt{ding20}); 
\href{https://emcee.readthedocs.io/en/stable/}{\texttt{emcee}} 
(\citealt{emcee13});
\href{https://dynesty.readthedocs.io/en/stable/}{\texttt{dynesty}} 
(\citealt{Speagle20});
\href{https://prospect.readthedocs.io/en/latest/}{\texttt{prospector} (\citealt{Johnson21})}
}

\bibliography{main}{}
\bibliographystyle{aasjournal}

\end{document}